# Observation of in-gap states in a two-dimensional $CrI_2$/$NbSe_2$ heterostructure

*Peigen Li,[§,‡,#] Jihai Zhang,[§,‡,#] Di Zhu,[§,‡] Cui-Qun Chen,[§,‡] Enkui Yi,[§,‡] Bing Shen,[§,‡] Yusheng Hou,[§,‡] Zhongbo Yan,[§,‡] Dao-Xin Yao,[§,‡] Donghui Guo,[§,‡] and Dingyong Zhong[§,‡,\*]*

[§]*School of Physics & Guangdong Provincial Key Laboratory of Magnetoelectric Physics and Devices, Sun Yat-sen University, 510275 Guangzhou, China*

[‡]*State Key Laboratory of Optoelectronic Materials and Technologies, Sun Yat-sen University, 510275 Guangzhou, China*

[#]*The authors equally contributed to this work.*

[\*]*Email: dyzhong@mail.sysu.edu.cn*

## Abstract

Low-dimensional magnetic structures coupled with superconductors are promising platforms for realizing Majorana zero modes, which have potential applications in topological quantum computing. Here, we report a two-dimensional (2D) magnetic-superconducting heterostructure consisting of single-layer chromium diiodide ($CrI_2$) on a niobium diselenide ($NbSe_2$) superconductor. Single-layer $CrI_2$ nanosheets, which hold antiferromagnetic (AFM) ground states by our first-principles calculations, were epitaxially grown on the layered $NbSe_2$ substrate. Using scanning tunneling microscopy/spectroscopy, we observed robust in-gap states spatially located at the edge of the nanosheets and defect-induced zero-energy peaks inside the $CrI_2$ nanosheets. Magnetic-flux vortices induced by an external field exhibit broken threefold rotational symmetry of pristine $NbSe_2$ superconductor, implying the efficient modulation of the interfacial superconducting states by the epitaxial $CrI_2$ layer. A phenomenological model suggests the existence of chiral edge states in a 2D AFM-superconducting hybrid system with an even Chern number, providing a qualitatively plausible understanding for our experimental observation.

There has been a great interest in designing materials emerging novel quantum phenomena, and one of the goals among the efforts is the search for Majorana zero modes (MZMs) in quantum materials. MZMs obey non-Abelian statistics, a remarkable property that is more resilient to local perturbation which may cause quantum decoherence.[1-3] Adiabatic braiding of MZMs is proposed to create a quantum-gate operation on the degenerate ground states[4], which is expected to be used for realizing fault-tolerant quantum computers[5, 6]. Topological superconductors with spin-triplet pairing are predicted to harbor MZMs, and yet remain scarce in the nature. Creative proposals such as the Kitaev chain[7] and the Fu-Kane[8] model, give inspiration to construct artificial topological superconductors to realize MZMs in hybrid systems. Evidence of MZMs has been observed in various heterostructure platforms, including one-dimensional (1D) Rashba nanowire/superconductor[9, 10], topological insulator/superconductor[11, 12] and magnet/superconductor[13-16]. The key components of these proposals are the strong spin-orbit coupling (SOC) or the spin-momentum locking combined with superconductivity to induce effective *p*-wave pairing.

Magnetic/superconducting hybrid systems used to construct topological superconductors have the advantage that the magnetic textures can spontaneously generate an intrinsic Zeeman field rather than requiring an external magnetic field. The groundbreaking experiment on a magnetic atomic Fe chain on a Pb(110) surface[17, 18] demonstrated that the signatures of MZMs can be observed at the ends of the chain, a hallmark of 1D topological superconductors. Signatures of MZMs have also been observed in other 1D magnetic atomic chain systems, such as Mn chains on Nb(110)[19] and Fe chains on Re(0001)[20]. On the other hand, a two-dimensional (2D) Shiba lattice[14, 15] consisting of magnetic atoms on an *s*-wave superconductor with a Rashba SOC was predicted to realize 2D chiral topological superconductivity with propagating chiral 1D Majorana edge states (MES). The experimental signatures of 1D chiral MES located at the boundary of magnetic atomic islands have been reported on the Pb/Co/Si(111)[21] and Fe/Re(0001)-O(2×1) systems[22]. However, the MES of these systems are sensitive to adsorption position of the magnetic atoms. Since the discovery of 2D intrinsic ferromagnetic (FM)[23-26] and antiferromagnetic (AFM)[27] materials, 2D van der Waals (vdW) magnetic materials have become excellent building blocks for magnetic/superconducting hybrid systems, holding the feasibility of fabricating high-quality heterostructures owing to the weak interfacial vdW interactions. Recently, Moiré-driven zero-bias states[28, 29] have been observed at the edges of FM monolayer $CrBr_3$ islands grown on $NbSe_2$. Furthermore, AFM/superconducting hybrid systems[30] have also been proposed to realize topological superconductors. A possible characteristics of MZMs were reported in a frustrated AFM $MnTe/Bi_2Te_3/Fe(Te,Se)$ heterostructure[31]. In addition, noncollinear spin textures such as magnetic skyrmions[32, 33] in proximity to *s*-wave superconductors were also predicted to induce topological superconducting states and stabilize MES even in the absence of SOC.

Here, we report the observation of in-gap edge states and defect-induced zero-bias conductance peaks (ZBCPs) in a vdW heterostructure consisting of a single-layer $CrI_2$ nanosheet grown on $NbSe_2$, a layered *s*-wave superconductor, by molecular beam epitaxy (MBE). Our first-principles calculations suggested that the monolayer $CrI_2$ is an AFM insulator. By means of high-resolution scanning tunneling microscopy/spectroscopy (STM/STS), we observed ZBCPs and in-gap edge states in the $CrI_2/NbSe_2$ hybrid system. Based on an effective Hamiltonian, we discussed the possibility of creating chiral edge states in a 2D AFM-superconducting hybrid system with an even Chern number.

Single-layer $CrI_2$ was grown on $NbSe_2$ surfaces with a submonolayer coverage by codeposition of chromium and iodine under ultrahigh vacuum conditions, as described in previous works[34, 35] (Figure 1a). Discrete flat $CrI_2$ nanosheets with an approximately round shape and a size in the range from several tens to a hundred nanometers were formed on the surface, as shown in Figure 1b. The regions between the $CrI_2$ nanosheets were covered by a nonstoichiometric Cr-I layer (see Figure S1). In the zoom-in STM topography (Figure 1c), there is a quasiperiodic Moiré pattern on the $CrI_2$ layer with the periodicity roughly 16 times than the primitive lattice, indicating an incommensurate relationship with the underlying $NbSe_2$ surface. Figure 1d is the fast

Fourier transform (FFT) image of Figure 1c, showing both the patterns corresponding to the Moiré superstructure and the topmost iodine atoms of $CrI_2$, marked by orange and blue circles, respectively. Inverse FFT of the blue-circled spots can reproduce the incommensurate Moiré pattern (Figure S2). It is known that Jahn-Teller effect in monolayer $CrI_2$[35, 36] induces lattice distortion and breaks the $C_3$ symmetry, which may promote the formation of the incommensurate Moiré pattern with the underlying $NbSe_2$. Figure 1e shows the atomic-resolved STM image of monolayer $CrI_2$ on the $NbSe_2$ surface. Compared with the case on Au(111) surfaces[35], the lattice constants of monolayer $CrI_2$ on $NbSe_2$ surfaces are slightly enlarged ($a_1$ = 4.16 Å, $a_2$ = 4.33 Å and $a_3$ = 4.31 Å), implying the existence of tensile strain. The large bias range of the differential conductance ($dI/dV$) spectrum on $CrI_2$ grown on $NbSe_2$ exhibits occupied and unoccupied states at −1.35 eV and 1.8 eV, respectively, corresponding to a gap of about 3.15 eV, demonstrating its insulating nature (Figure S3). The small bias range of the $dI/dV$ spectrum on $CrI_2$ is similar to that of bare $NbSe_2$.

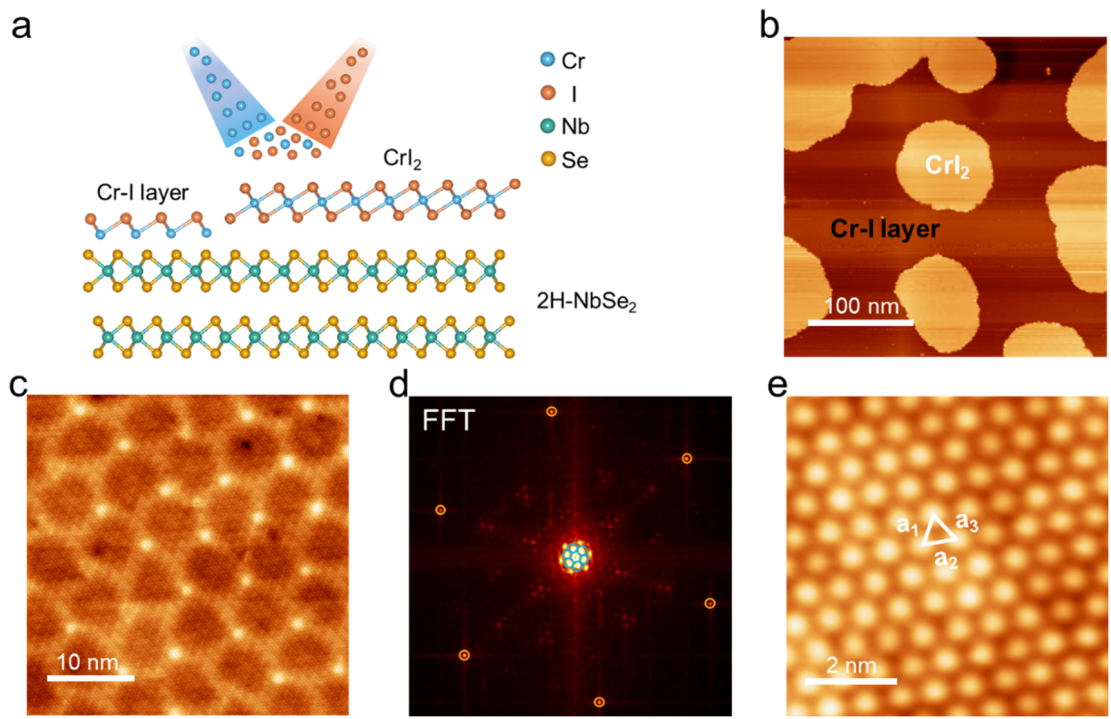

**Figure 1** Single-layer $CrI_2$ nanosheets on $NbSe_2$. (a) Schematic illustration of the growth of 2D $CrI_2/NbSe_2$ heterostructures. (b) Large-scale STM topography showing the round-shaped single-layer $CrI_2$ nanosheets ($V_s$ = 2.0 V, $I_t$ = 10 pA). (c) Zoom-in STM image of single-layer $CrI_2$ showing

the incommensurate Moiré pattern ($V_s$ = 1.5 V, $I_t$ = 150 pA). (d) FFT pattern of (c) indicating the coexistence of the primitive $CrI_2$ lattice and the approximate (16×16) Moiré superstructure. (e) Atomic-resolution STM image showing the distorted closed-packed structure of iodine atoms at the surface of $CrI_2$ nanosheets ($V_s$ = 1.0 V, $I_t$ = 200 pA).

The *dI/dV* spectra were collected from the $CrI_2$/$NbSe_2$ heterostructure at a temperature of 350 mK. As shown in Figure 2a and 2b, the *dI/dV* spectra from the internal region of the $CrI_2$ island exhibit a typical U-shaped superconducting gap with a pair of coherent peaks located at ±1.2 mV, which can be fitted by the *s*-wave BCS-type scenario (Figure S4), similar to that of bare $NbSe_2$. Interestingly, at the edge of $CrI_2$ nanosheets, besides similar superconducting coherent peaks, a distinguishable non-zero conductance signal exists inside the superconducting gap. To further reveal the spatial distribution of the in-gap states, we measured a series of *dI/dV* spectra from the edge to interior of the $CrI_2$ island (Figure 2c, also see Figure S5). It was found that the in-gap states are strongest at the edge and become weaker and finally disappear at about 9 nm away from the edge, indicating the spatially localized nature. To quantitatively characterize the spatial distribution of the in-gap edge states, we extracted the integral area of the *dI/dV* spectra ranging from −0.25 mV to 0.25 mV as a function of the distance away from the island edge (Figure S6). The results reveal that the spatial extension length of the in-gap edge states is about 6 ~ 9 nm. In some cases, additional pair of Yu-Shiba-Rusinov (YSR) states was observed (Figure S7), which originate from the interactions between the defect-induced local magnetic moments and Cooper pairs. Nevertheless, the in-gap edge states are preserved, implying the robustness against defects. Notably, a non-stoichiometric Cr-I layer surrounding the $CrI_2$ islands is ubiquitous in our sample, the effect of which on the electronic states of $CrI_2$ islands should be considered. As shown in Figure S8, the Cr-I layer exhibits a V-shaped non-zero conductance signal inside the superconducting gap. Additionally, the line profile of *dI/dV* spectra display a uniform V-shaped feature in the Cr-I layer region (Figure S8). To rule out the possibility that the in-gap edge states on $CrI_2$ islands are induced by the Cr-I layer, we prepared a sample with isolated Cr-I islands grown on NbSe$_2$ (Figure S9). There exists a clean boundary between Cr-I layer and NbSe$_2$ surface, therefore we can investigate the spatial distribution of the V-shaped in-gap states of Cr-I layer. Similar V-shaped feature appears in the Cr-I layer region, which is weakened at the edge of the Cr-I layer. Such in-gap states are rapidly decayed with a length about 4 nm at the NbSe$_2$ side, indicating the localized feature of the in-gap states induced by Cr-I layer. On the contrary, the significantly larger spatial distribution of the in-gap states at the CrI$_2$ edge supports that these states originate from the intrinsic interactions between single-layer CrI$_2$ and NbSe$_2$. Unlike the FM CrBr$_3$/NbSe$_2$ system[28] exhibiting ZBCPs at the edges, the edge states observed here show a continuum feature dispersing in the whole superconducting gap, with a faint intensity at zero bias.

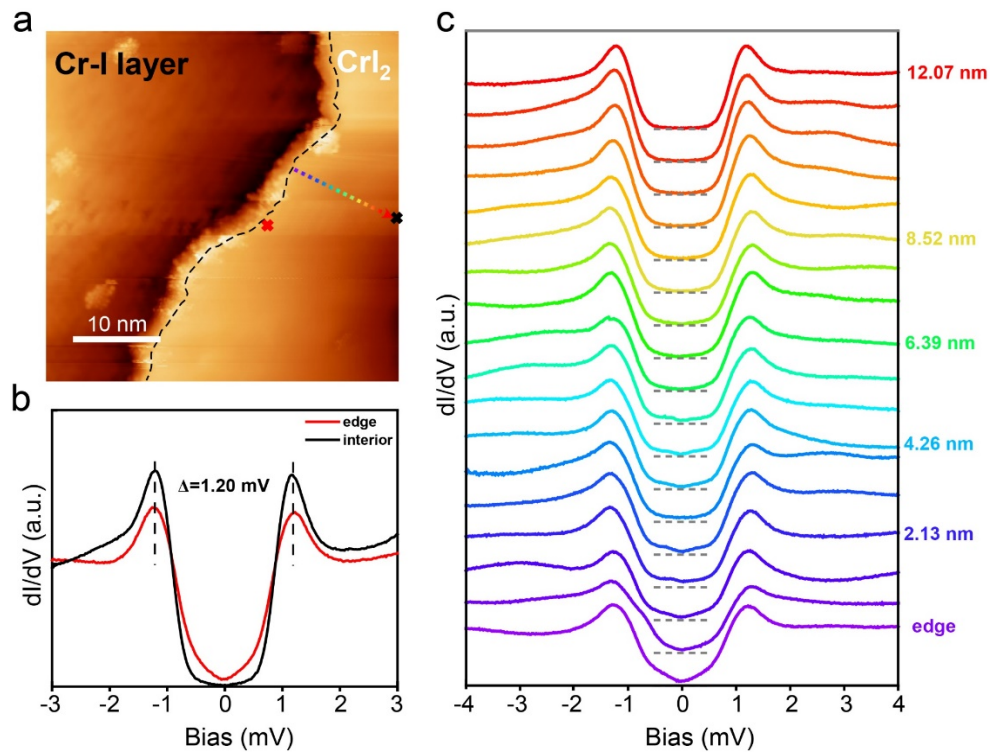

**Figure 2** *dI/dV* spectra of single-layer CrI$_2$/NbSe$_2$ heterostructure. (a) STM topography by the edge of a CrI$_2$ island ($V_s = -1.0$ V, $I_t = 40$ pA). The edge of CrI$_2$ is traced by the black dashed line. (b) *dI/dV* spectra acquired from the edge and interior of the CrI$_2$ island (marked by crosses in (a)), respectively. (c) A series of *dI/dV* spectra from the CrI$_2$ edge to the inside (marked by colored arrow in (a)). The distance away from the CrI$_2$ edge is labeled in (c). The gray dashed lines represent the zero reference values.

Figure 3a shows the *dI/dV* mapping of a single-layer CrI$_2$ island at a bias of 0.25 mV (*dI/dV* mapping for other energies see Figure S10). At the edge (marked by the white dotted line), there are distinct non-zero conductance signals ascribed to the in-gap states, which is continuously distributed around the island with the intensity decayed inside the island. Interestingly, although the conductance is mostly close to zero in the internal region of the island, there exist some individual spots exhibiting obviously intensive conductance signals. (marked by red arrows in Figure 3a). The atomically resolved STM image of such region (Figure 3b) shows irregular contrast, implying the existence of structural defects. Specifically, there is relatively brighter hexagon containing 7 topmost atoms (marked by yellow circle), probably originating from the substitution of a surface iodine atom by a chromium atom (marked by a green cross). Moreover, the nearby slightly darker contrasts (marked as black triangles) centered at the hollow sites may be ascribed to subsurface vacancies. Irreversible structural transformation induced by STM-tip was observed, resulting in a dark hole defect (white circle) and a brighter spot defect (yellow circle), as shown in Figure 3c. We measured the *dI/dV* spectra before and after transformation. As shown in Figure 3d, the *dI/dV* spectrum before transformation exhibits a pair of in-gap peaks located at ± 0.23 mV, a prominent feature of YSR states[37-39]. In contrast, the *dI/dV* spectrum after transformation (at the same position) shows a strong ZBCP. The half-peak width of the ZBCP is about 0.28 meV, close to the energy resolution of our STM (see Figure S11). Figure 3e and 3f show the STS results across this defect along two orthogonal directions. Sharp ZBCP signals can be seen in the vicinity of the defect, with strongest intensity at the center of the bright spot defect (yellow circle). As away from the center, the intensity is gradually weakened and no splitting occurred. Figure 3g shows the zero-bias *dI/dV* mapping at the defect region, indicating that the ZBCP is mainly distributed at the bright spot defect with a spatial extension of about 4 nm.

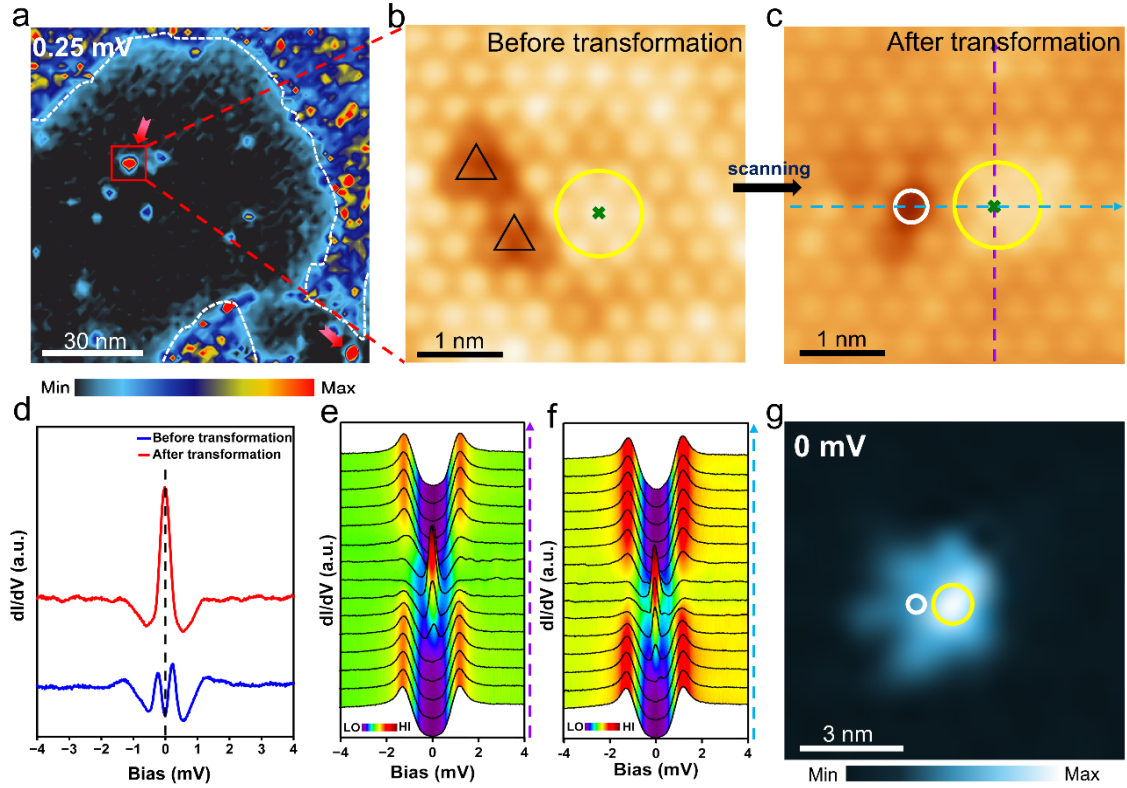

**Figure 3** Edge states and defect-induced zero-energy peak. (a) *dI/dV* mapping with E = 0.25 mV at B = 0 T. The $CrI_2$ edge is traced by the white dashed line. Red arrows denote the regions with strong conductance signal. (b) STM image of the red rectangular region in (a) ($V_s$ = −0.7 V, $I_t$ = 100 pA). The black triangles denote three nearest-neighbor topmost iodine atoms exhibiting darker contrast. (c) STM image at the same region as (b) after structural transformation ($V_s$ = −0.7 V, $I_t$ = 100 pA). There are a brighter bulge and an iodine vacancy, denoted by yellow and white circles, respectively. (d) The *dI/dV* spectra before and after structural transformation at the position marked by the green crosses in (b) and (c). (e, f) Series of *dI/dV* spectra measured across the defect region along the purple (e) and blue (f) dashed arrows as shown in (c). (g) The spatial distribution of zero-energy peaks. The bulge and iodine vacancy are marked similar as (c).

Defect-induced ZBCPs have been previously reported in several superconducting systems[31, 40-44]. For example, interstitial magnetic Fe impurities[40] and atomic line defects[41] at the surface of superconducting Fe(Te,Se) exhibit robust ZBCPs, which were considered as the characteristics of MZMs. Defect-induced ZBCPs were also observed in a trilayer heterostructure of $MnTe/Bi_2Te_3/Fe(Te,Se)$,[31, 45] which were ascribed to the magnetic Mn-Bi antisite defects sandwiched between the $Bi_2Te_3$ and MnTe layers. Similar transformation between YSR states and ZBCPs was also observed in the case of Fe adatoms at the surface of $FeTe_{0.55}Se_{0.45}$[42], which was initiated by the modulation of the adsorbate-surface interactions. In our experiments, it is reasonable to infer that the morphology evolution around the $CrI_2$ defects is accompanied with the change of the coupling strength between the local magnetic impurity and superconducting substrate, resulting in the transformation of the in-gap states.

$NbSe_2$ substrate is a type-II superconductor[46]. By applying an out-of-plane magnetic field to the $CrI_2$/$NbSe_2$ heterostructure, we have observed the Abrikosov vortex on the heterostructure surface. Figure 4a shows a typical zero-bias $dI/dV$ mapping at the $CrI_2$ surface with a vortex under an out-of-plane magnetic field of 0.1 T. The $dI/dV$ spectrum at the vortex center on the $CrI_2$ surface exhibits a sharp ZBCP (inset of Figure 4a). To further dissect the spatial evolution of the ZBCP, we measured the $dI/dV$ spectra across the vortex center along the cut 1 and 2 directions (orange and blue dashed lines), plotted in Figure 4b and 4c. The ZBCPs are held about 5 nm around the vortex center, and then split into two symmetric peaks inside the superconducting gap as the tip moves away from the vortex center. Meanwhile, we found that the energy of the splitting peaks approximately linearly increases with distance (black dashed line in Figure 4b and 4c). Limited by the energy resolution (~ 0.23 meV), it is difficult to judge whether the ZBCPs are persistent away from the vortex center, coexisting with the trivial Caroli-de Gennes-Matricon (CdGM) bound states[46-49]. However, earlier work[11, 50] reported that the MZM can be maintained near 20 nm near the vortex center. Therefore, the contribution of ZBCPs near the vortex center in our case may be not derived from MZM but from the CdGM bound states.

With a larger magnetic field of 0.65 T, we found that the spatial distribution of the vortices is elongated along the cut 2 direction, exhibiting an elliptical shape (Figure 4d and Figure S12). The evolution of the $dI/dV$ spectra across the vortex along the cut 1 and 2 directions are plotted in Figure 4e and 4f, respectively. For the cut 1 direction, the splitting behavior of ZBCPs (Figure 4e) is symmetric with respect to the vortex center, similar to the splitting characteristics in 0.1 T field. For the cut 2 direction, however, the splitting behavior (Figure 4f) exhibits an anisotropic feature. Here, at an external magnetic field of 0.65 T, the average distance between neighboring vortices is about 60 nm, much larger than the in-plane superconducting coherence length[51] of $NbSe_2$. Thus, the interactions between neighboring vortices, as one of the possible factors affecting the vortex shape, is ruled out. Anisotropic Fermi surface[52] also affects the shape of vortex. In the case of clean $NbSe_2$, which possesses a Fermi surface with sixfold symmetry, leads to a sixfold star-shape vortex[53]. When monolayer $CrI_2$ is epitaxially grown on the $NbSe_2$ surface, the coupled interfacial interactions may affect the Fermi surface of the underlying $NbSe_2$ and effectively modulate the interfacial superconducting state, breaking the pristine $C_3$ rotational symmetry.

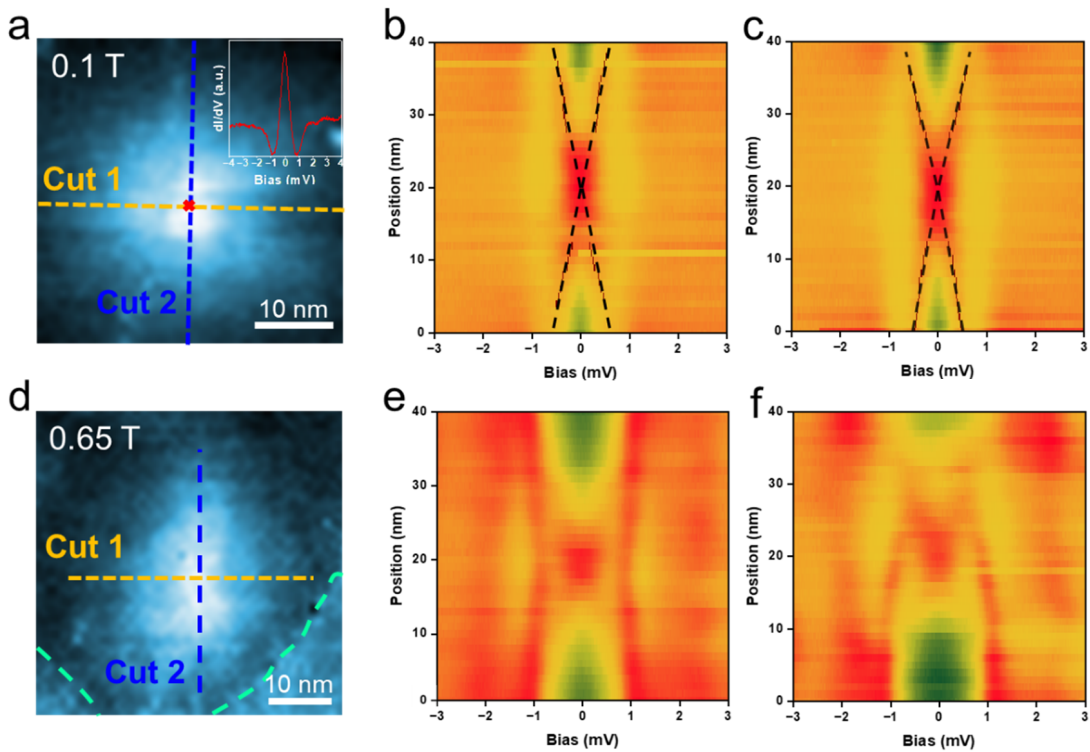

**Figure 4** Abrikosov vortex in $CrI_2/NbSe_2$ heterostructure under an out-of-plane magnetic field. (a) Zero-bias *dI/dV* mapping of a vortex with B = 0.1 T. Inset, *dI/dV* curve at the vortex center showing a sharp ZBCP. (b, c) False-colour images of a series of *dI/dV* spectra measured along cut 1 (b) and cut 2 (c) marked in (a). Peak splitting is denoted by black dashed lines. (d) Zero-bias *dI/dV* mapping of a vortex with B = 0.65 T. The edge of $CrI_2$ is traced by the green dashed line. (e, f) False-colour images of a series of *dI/dV* spectra measured along cut 1 (f) and cut 2 (g) marked in (d).

Our density functional theory (DFT) calculations reveal a stripe AFM ground state of $CrI_2$ both in freestanding monolayer and monolayer on a $NbSe_2$ substrate. The vdW gap between $CrI_2$ and $NbSe_2$ is 3.28 Å. Compared to the lattice constant in previous study[35], the lattice of $CrI_2$ in $CrI_2$/$NbSe_2$ heterostructures exhibits a tensile strain (~4%). To investigate the effect of strain, we further performed calculations on the magnetic ground state of monolayer $CrI_2$ under strain. We found that under compression strain and minor tensile strain (from 0 ~ 5%), $CrI_2$ persists the AFM ground state, whereas FM becomes the ground state when the tensile strain is larger than 7%, as shown in Figure S13.

The in-gap edge states and stable ZBCPs at defects provide distinct evidence of topological superconducting states in $CrI_2$/$NbSe_2$ heterostructures. A phenomenological model was built from the view point of symmetry to give an effective description to the 2D interface between single-layer antiferromagnetic $CrI_2$ and superconducting $NbSe_2$. In this model (Figure 5a), as magnetism generally suppresses spin-singlet pairing and Rashba SOC favors a mixture of spin-singlet and spin-triplet pairings[54], we took a chiral *p*-wave pairing component into account (detailed model descriptions see Section 12 of SM). We found that chiral topological superconductors with chiral MES can be realized when the *p*-wave component dominates in the mixed-parity pairing. However, because the zero net magnetization of the AFM order and SOC constrain the number of Fermi surfaces to be even, the chiral topological superconducting phase must have an even Chern number[55]. To be specific, we found that the model gives rise to a chiral topological superconductor with Chern number $C = 2$, as shown in Figure 5b. Under open boundary conditions, we found the existence of two branches of chiral MES on each edge, in consistency with the so-called bulk-boundary correspondence. By calculating the density of states under open (with chiral MES) and periodic (no chiral MES as there is no edge) boundary conditions (Figure 5c), we did find that the chiral MES contribute finite density of states in the superconducting gap, which is qualitatively in agreement with the experimental observation presented in Figure 2b. In addition, simulated spatial distributions of edge states are mainly concentrated at the edges of 2D islands. Nevertheless, a chiral topological superconductor with Chern number $C = 2$ results in the absence of stable ZBCP at the vortices. According to the bulk-vortex correspondence[56], a vortex in a chiral topological superconductor with $C = 2$ will carry two MZMs. However, because the vortex MZMs in general (without additional symmetry protection) obey a $Z_2$ classification[56], two MZMs will hybridize and split to finite energy.

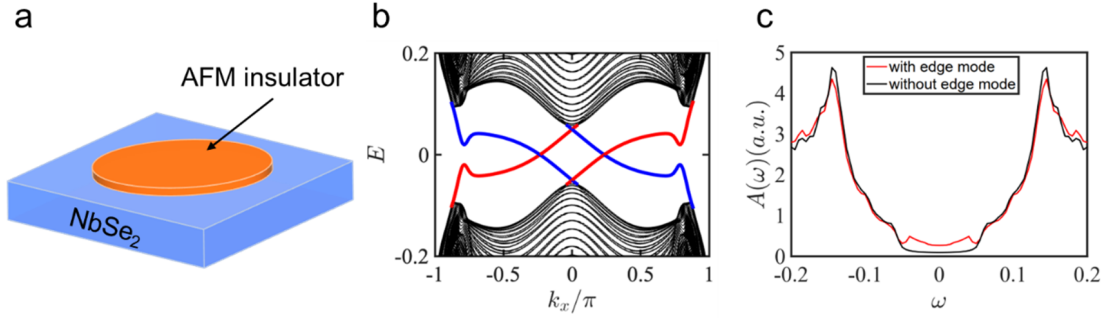

**Figure 5** Modeling of topological superconductors about an AFM insulator on *s*-wave superconductor. (a) Schematic of the model. (b) Chiral Majorana edge states for a chiral topological superconductor with Chern number equal to two. The two red and two blue lines represent two branches of chiral Majorana modes on the upper and lower *y*-normal edges, respectively. (c) The density of states (in arbitrary unit) for the same set of parameters as in (b). The red (black) solid line corresponds to system with open (periodic) boundary conditions in the *y* direction (100 unit cells) and periodic boundary conditions in the *x* direction. Chosen parameters of the Hamiltonian are: t=1, $\lambda_R$=0.1, μ=1.7, $M_z$=0.05, $\Delta_s$=0.05, and $\Delta_p$=0.12 (details of the Hamiltonian see Section 8 of SM). In calculating the density of states, we have replaced the δ function by a Lorentz function with a broadening factor Γ=0.004, so the density of states do not reach exactly zero in the superconducting gap.

In summary, we successfully prepared a 2D magnetic-superconducting hybrid heterostructure consisting of single-layer $CrI_2$ on $NbSe_2$ substrates by MBE. STS measurements revealed that there are in-gap states near the $CrI_2$ edge and these in-gap states extend to about 9 nm inside the $CrI_2$. Stable ZBCPs have been observed on individual defects of $CrI_2$, with a spatial scale up to 4 nm. When a magnetic field is applied, vortices appear at the surface of single-layer $CrI_2$ by superconducting proximity effect, and the splitting characteristics of the ZBCPs at the vortices are anisotropic with increasing magnetic field. A phenomenological model was proposed, suggesting that a chiral topological superconductor with an even Chern number can provide a qualitatively plausible and consistent interpretation of the experimental observations of in-gap edge states and the behaviors of the ZBCP at the vortices. Our work provides an appealing platform to explore possible topological superconducting phases and MZMs based on 2D magnetic/superconducting hybrid systems.

**Methods**

MBE Sample Growth: Single-layer $CrI_2$ films were grown on a freshly cleaved $NbSe_2$ substrate by MBE under ultra-high vacuum (UHV) condition (base vacuum $1\times10^{-10}$ mbar). Iodine source was obtained by thermal decomposition of $CrI_3$ powder (purity 90 %) at 230 °C. Chromium powder (purity 99.996 %, Alfa Aesar) was evaporated from a Knudsen cell at 1025 °C. Single-layer $CrI_2$ films were grown using a two-step method, i.e., codeposition of Cr and I on the $NbSe_2$ surface for 10 min at room temperature followed by annealing to 230 °C.

STM and STS Measurements: After sample preparation, the sample was inserted into a low-temperature STM (Unisoku USM-1300) housed in the same UHV system. Typical STM images were taken at 5.2 K and STS measurements were taken at 350 mK. All STM measurements were performed with constant current mode using a PtIr tip. The *dI/dV* spectra were measured by using the lock-in technique with a reference signal at 983 Hz. The modulation amplitudes were set as 20 ~ 30 μV. The STM images in the article were processed using the WSxM software[57].

Calculation details: Density functional theory (DFT) calculations were performed using the generalized gradient approximation (GGA) with Perdew–Burke–Ernzerhof (PBE) form[58] for the exchange−correlation potential with a plane-wave basis and the projected-augmented wave (PAW) method[59] as implanted in the Vienna ab initio Simulation Package (VASP)[60]. The energy cutoff for the plane wave was set to 500 eV and an effective Hubbard U for the 3$d$ electrons of Cr cations was chosen as 4 eV[35]. The vdW correction in the form of the semiempirical DFT-D3 method[61] was adopted. The lattice parameter of CrI$_2$ was fixed using the experimentally measured lattice constant and the positions of all atoms were fully relaxed until the force on each atom is less than 0.01 eV/Å. The heterostructure of CrI$_2$/NbSe$_2$ consists of a 3×6×1 supercell of CrI$_2$ and a $\sqrt{13} \times 2\sqrt{13}$ supercell of NbSe$_2$. A 4×2×1 k-point mesh was employed and the energy convergence criteria was 10$^{-6}$ eV.

**Supporting Materials**

The Supporting Materials are available and free of charge at xxx.

Atomic structure of the Cr-I layer; The Moiré superstructure of single-layer CrI$_2$ on the NbSe$_2$ surface; $dI/dV$ spectra of single-layer CrI$_2$ on NbSe$_2$; BCS fitting of the superconducting gap of the interior of CrI$_2$ island; Evolution characteristics of in-gap states; Effect of the defect on in-gap edge states; $dI/dV$ spectra of Cr-I layer; Spatial distribution of in-gap states with increasing energy; Calibration of energy resolution at T = 350 mK; Vortices on the surface of CrI$_2$ under a 0.65 T out-of-plane magnetic field; Magnetic ground state of monolayer CrI$_2$; Model description.

**Author contributions**

P.G.L., J.H.Z. and D.Y.Z. conceived the research and designed the experiments. P.G.L. and J.H.Z. performed the STM/S experiments. D.Z. and Z.B.Y. proposed the theoretical model and performed the numerical simulation. C.-Q.C., Y.S.H. and D.-X.Y. performed the first principles calculation. E.K.Y. and B.S. provided the NbSe$_2$ single crystal. P.G.L., J.H.Z., D.Z., C.-Q.C., Y.S.H., Z.B.Y., D.-X.Y., D.H.G. and D.Y.Z. did the data analysis and discussed the results. P.G.L., Z.D., C.-Q.C., Z.B.Y., D.-X.Y. and D.Y.Z. wrote the paper with input from all co-authors. P.G.L. and J.H.Z. equally contributed to the work.

**Notes**

The authors declared no competing financial interest.

**Acknowledgment**

The work was financially supported by the National Natural Science Foundation of China (92165204, 11974431 and 12174455), the National Key Research and Development Program of China (2022YFA1402802 and 2018YFA0306001), Guangdong Major Project of Basic and Applied Basic Research (2021B0301030002), the Shenzhen Institute for Quantum Science and Engineering (SIQSE202102), and the Guangdong Provincial Key Laboratory of Magnetoelectric Physics and Devices (2022B1212010008).

TOC:

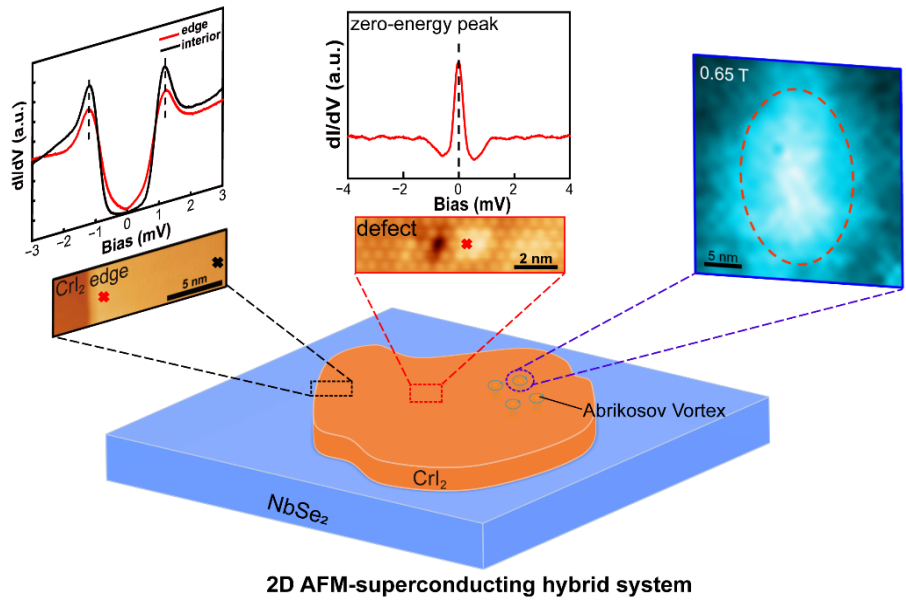

**2D AFM-superconducting hybrid system**